\documentclass[10pt]{article}
\usepackage{multirow}
\usepackage[dvips]{graphicx}

\usepackage[psamsfonts]{amssymb}
\usepackage{amsxtra}
\usepackage{threeparttable}
\usepackage{amsmath}
\usepackage{amssymb}

\usepackage{graphicx}

\usepackage{cite}

\usepackage{color}

\usepackage{setspace}
\usepackage{float}

\usepackage{url}

\usepackage[labelfont=bf,labelsep=period,justification=raggedright]{caption}

\makeatletter
\renewcommand{\@biblabel}[1]{\quad#1.}
\makeatother

\date{April 18, 2013}

\begin{document}

\begin{flushleft}
{\Large
\textbf{Multi--command Chest Tactile Brain Computer Interface for Small Vehicle Robot Navigation}
}

Hiromu Mori$^1$, 
Shoji Makino$^1$, 
and Tomasz M. Rutkowski$^{1,2,}$\footnote{The corresponding author.}$^,$\footnote{The final publication is available at http://link.springer.com/}
\\
\bf{1} Life Science Center of TARA, University of Tsukuba, Tsukuba, Japan\\
\bf{2} RIKEN Brain Science Institute, Wako-shi, Japan\\
$\ast$ E-mail: tomek@tara.tsukuba.ac.jp\\
http://about.bci-lab.info/
\end{flushleft}

\section*{Abstract}

The presented study explores the extent to which tactile stimuli delivered to five chest positions of a healthy user can serve as a platform for a brain computer interface (BCI) that could be used in an interactive application such as robotic vehicle operation. The five chest locations are used to evoke tactile brain potential responses, thus defining a tactile brain computer interface (tBCI). Experimental results with five subjects performing online tBCI provide a validation of the chest location tBCI paradigm, while the feasibility of the concept is illuminated through information-transfer rates. Additionally an offline classification improvement with a linear SVM classifier is presented through the case study.

\noindent{\bf Keywords:} tactile BCI, P300, robotic vehicle interface, EEG, neurotechnology

\section{Introduction}

Contemporary BCIs are typically based on mental visual and motor imagery paradigms, which require extensive user training and good eyesight from the users~\cite{bciBOOKwolpaw}. Recently alternative solutions have been proposed to make use of spatial auditory~\cite{iwpash2009tomek,aBCIbirbaumer2010,bciSPATIALaudio2010} or tactile (somatosensory) modalities~\cite{sssrBCI2006,tactileBCIwaste2010,HBCIscis2012hiromuANDtomek,JNEtactileBCI2012,tomekHAID2013,tactileAUDIOvisualBCIcompare2013} to enhance brain-computer interface comfort and increase the information transfer rate (ITR) achieved by users. The concept reported in this paper further extends the previously reported by the authors in~\cite{tomekHAID2013} brain somatosensory (tactile) channel to allow targeting of the tactile sensory domain for the operation of robotic equipment such as personal vehicles, life support systems, etc. The rationale behind the use of the tactile channel is that it is usually far less loaded than auditory or even visual channels in interfacing applications.  

\begin{figure}
	\begin{center}
	\includegraphics[width=0.8\linewidth]{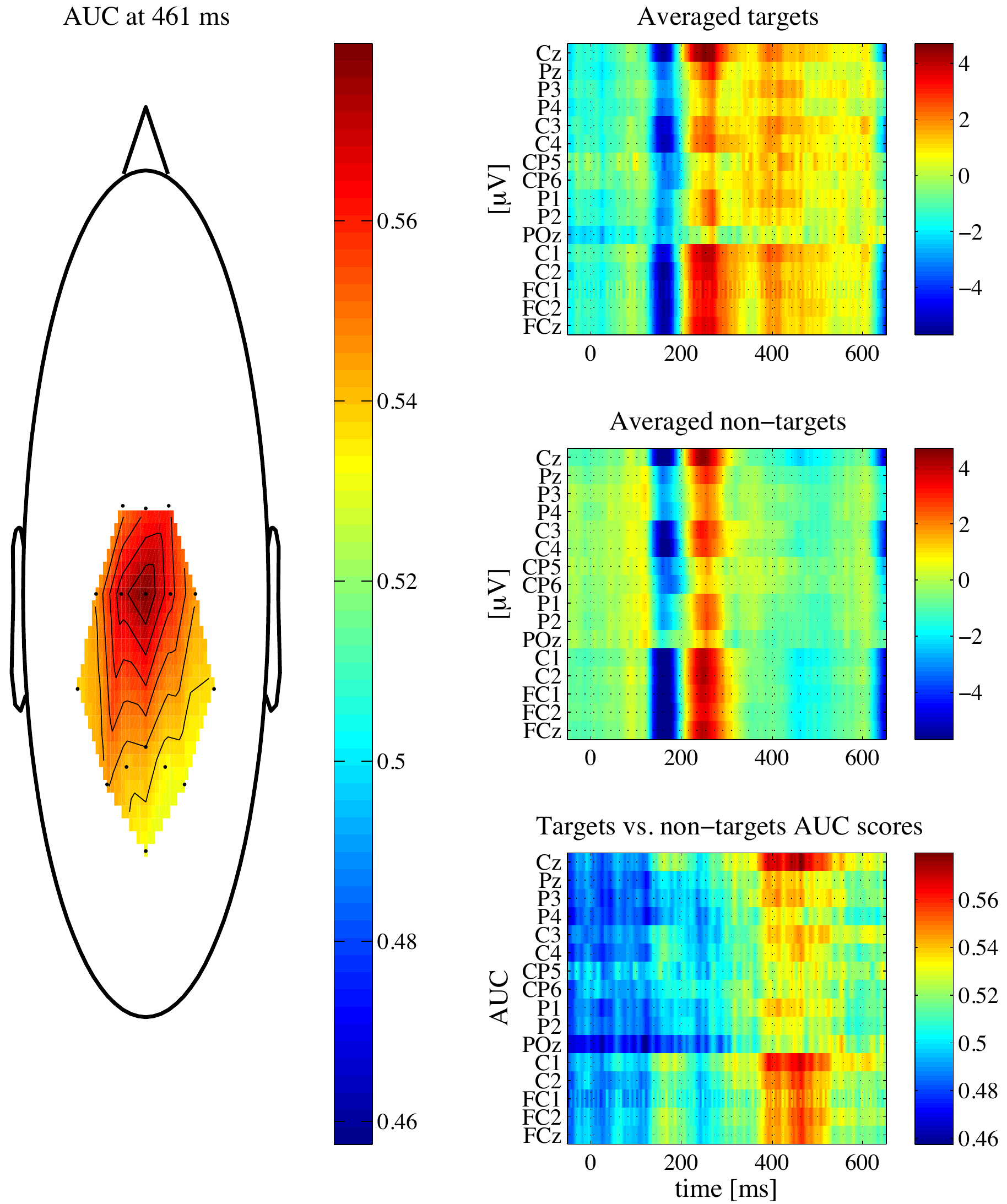}
	\end{center}
	%\vspace{-1cm}
	\caption{All the five subjects grand mean averaged results of the chest stimulation EEG experiment. The left panel presents the head topographic plot of the $target$ versus $non-target$ area under the curve (AUC), a measure commonly used in machine learning intra--class discriminative analysis. ($AUC > 0.5$ is usually assumed to be confirmation of feature separability~\cite{book:pattRECO}). The top right panel presents the largest difference as obtained from the data displayed in the bottom panel. The topographic plot also depicts the electrode positions. The fact that all the electrodes received similar AUC values (red) supports the initial electrode placement. The second panel from the top presents averaged SEP responses to the $target$ stimuli (note the clear P300 response in the range of $350-600$~ms). The third panel presents averaged SEP responses to the $non-target$ stimuli (no P300 observed). Finally, the bottom panel presents the AUC of $target$ versus $non-target$ responses (again, P300 could easily be identified).}\label{fig:allTOPO}
\end{figure}

\begin{figure}
	\begin{center}
	\includegraphics[width=0.8\linewidth]{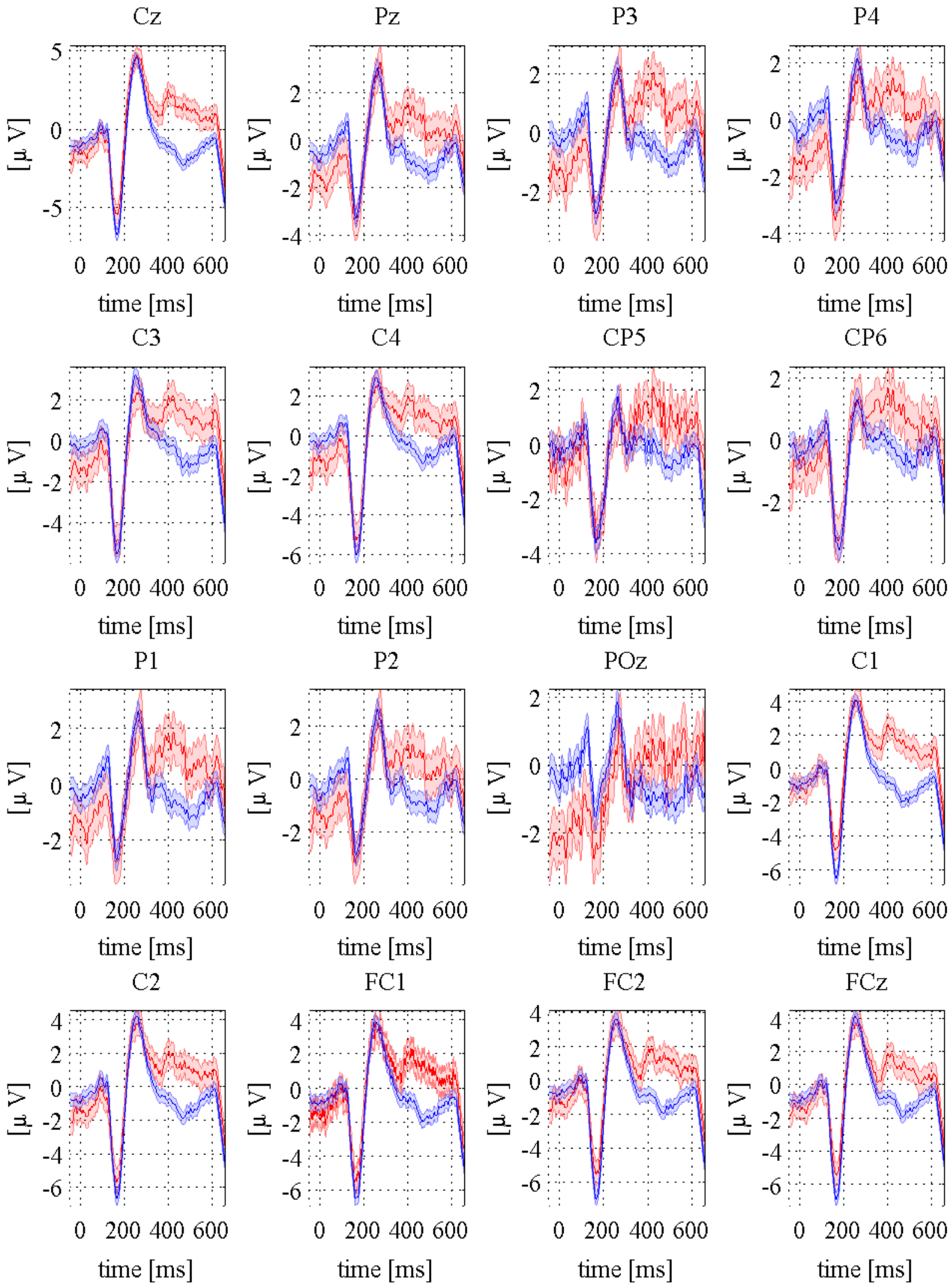}
	\end{center}
	\caption{All five subjects grand mean averaged results for each electrode plotted separately. The red and blue lines depict targets and non--targets respectively together with standard error bars. The P300 related responses could be observed in the $350-650$~ms latencies.}\label{fig:allERP}
\end{figure}

\begin{figure}
	\begin{center}
	\includegraphics[width=\linewidth, clip]{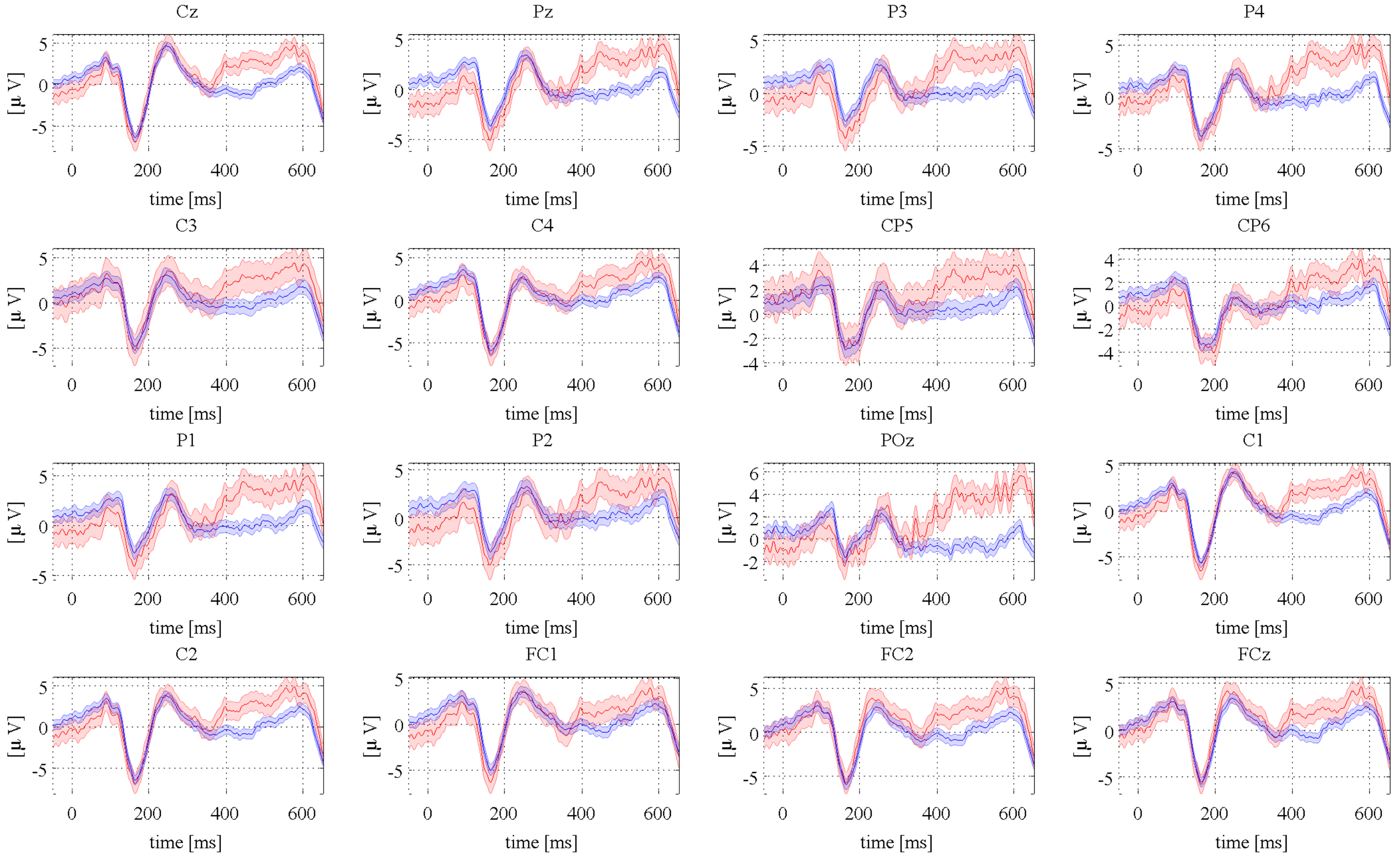}
	\end{center}
	\caption{Averaged results for each electrode separately of the subject~$\#3$ for whom $P300$ response was dominating leading for the best latency used in classification. The red and blue lines depict targets and non--targets respectively together with standard error bars.}\label{fig:s3ERP}
\end{figure}

\begin{figure}
	\begin{center}
	\includegraphics[width=\linewidth]{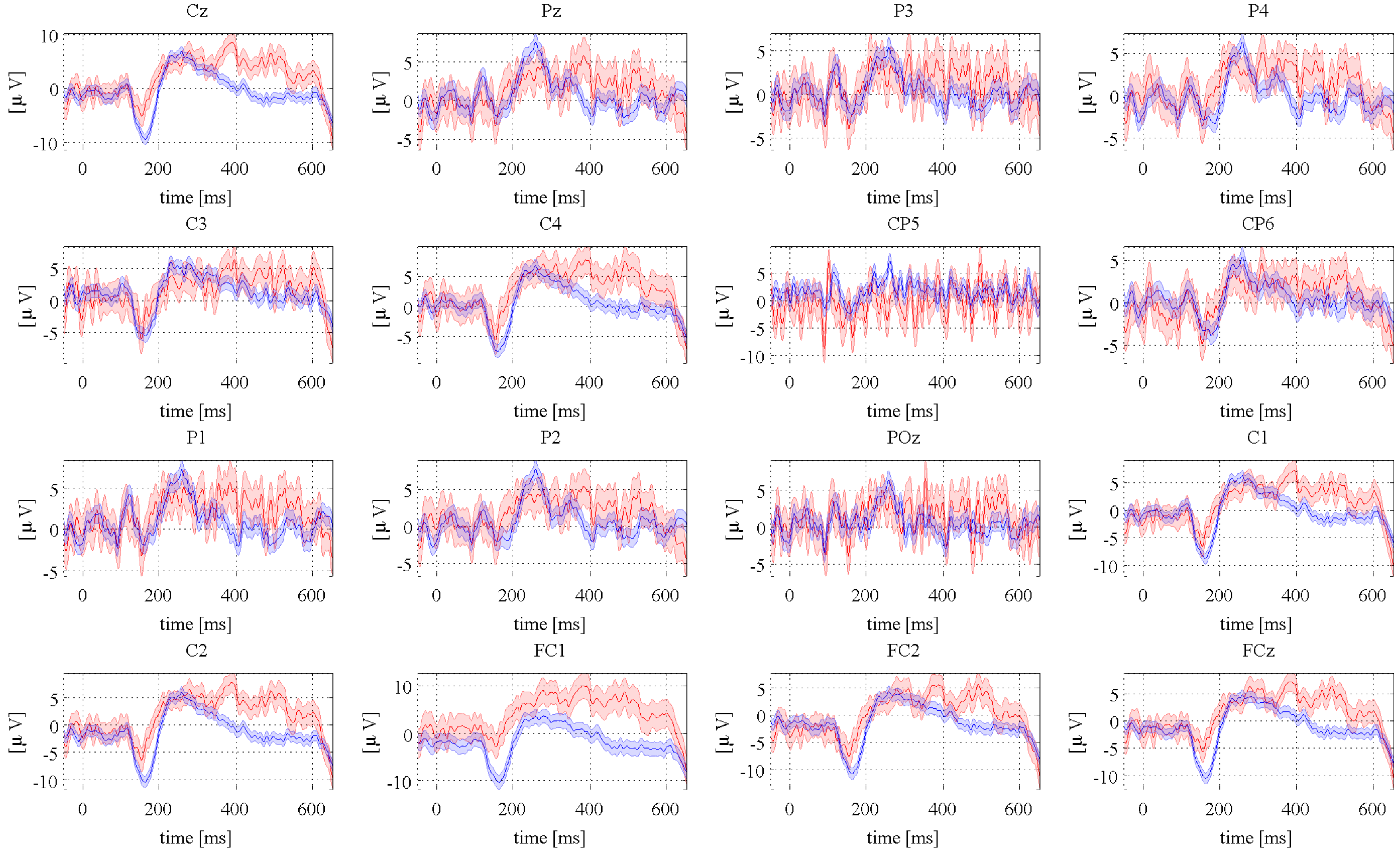}
	\end{center}
	\caption{Averaged results for each electrode separately of the subject~$\#2$ for whom $N200$ response was significant leading for improved early latency used in classification. Observe the very short latency around $200$ms where the standard error bars don't overlap. The red and blue lines depict targets and non--targets respectively together with standard error bars.}\label{fig:s2ERP}
\end{figure}

The first report~\cite{sssrBCI2006} of the successful employment of steady-state somatosensory responses 
to create a BCI targeted a low frequency vibrotactile stimulus in the range of $20-31$~Hz to evoke the subjects' attentional modulation, which was then used to define interfacing commands. A more recent report~\cite{JNEtactileBCI2012} proposed using a Braille stimulator with $100$~ms static push stimulus delivered to each of six fingers to evoke a somatosensory evoked potential (SEP) response and the following P300.
The P300 response is a positive electroencephalogram event-related potential (ERP) deflection starting at around $300$~ms and lasting for $200-300$~ms after an expected stimulus in a random series of distractors (the so-called oddball EEG experimental paradigm)~\cite{book:eeg}. 
Examples of averaged P300 response are depicted with red lines with standard errors in Figures~\ref{fig:allTOPO},~\ref{fig:allERP},~\ref{fig:s3ERP},~and~\ref{fig:s2ERP}.
The P300 responses are commonly used in BCI approaches and are considered to be the most reliable ERPs~\cite{bci2000book,bciBOOKwolpaw} with even beginner subjects. The results in~\cite{JNEtactileBCI2012} indicated that the experiments achieved information transfer rates of $7.8$~bit/min on average and $27$~bit/min for the best subject. 
A very recent report~\cite{tactileAUDIOvisualBCIcompare2013} additionally confirmed superiority of the tactile BCI (tBCI) in comparison with visual and auditory modalities tested with a locked--in syndrome (LIS) subject~\cite{alsTLSdiagnosis1966}.

This paper reports improvement of our previously reported finger stimulus tBCI~\cite{tomekHAID2013} based on P300 responses evoked by tactile stimulus delivered via vibrotactile transducers attached to five positions on the subject's chest this time. The proposal is similar to the previously reported waist positions based tBCI reported in~\cite{tactileBCIwaste2010} with a difference that we propose to use the chest area simulation which simplifies a vehicular robot operation in comparison to our previous hand- and head-stimulus--based tBCI solutions reported in~\cite{tomekHAID2013,HiromuBCImeeting2013}.

The rest of the paper is organized as follows. The next section introduces the materials and methods used in the study. It also outlines the experiments conducted. The results obtained in electroencephalogram online and offline experiments with five BCI subjects are then discussed. Finally, conclusions are formulated and directions for future research are outlined. 

\section{Materials and Methods}

Five volunteer male BCI subjects participated in the experiments. The subjects' mean age was $26$, with a standard deviation of $9.5$. All the experiments were performed at the Life Science Center of TARA, University of Tsukuba, Japan. 
The online (real-time) EEG tBCI paradigm experiments were conducted in accordance with the \emph{WMA Declaration of Helsinki - Ethical Principles for Medical Research Involving Human Subjects}. 

\subsection{Tactile Stimulus Generation}

\begin{figure}[!h]
	\begin{center}
	\includegraphics[width=0.5\linewidth]{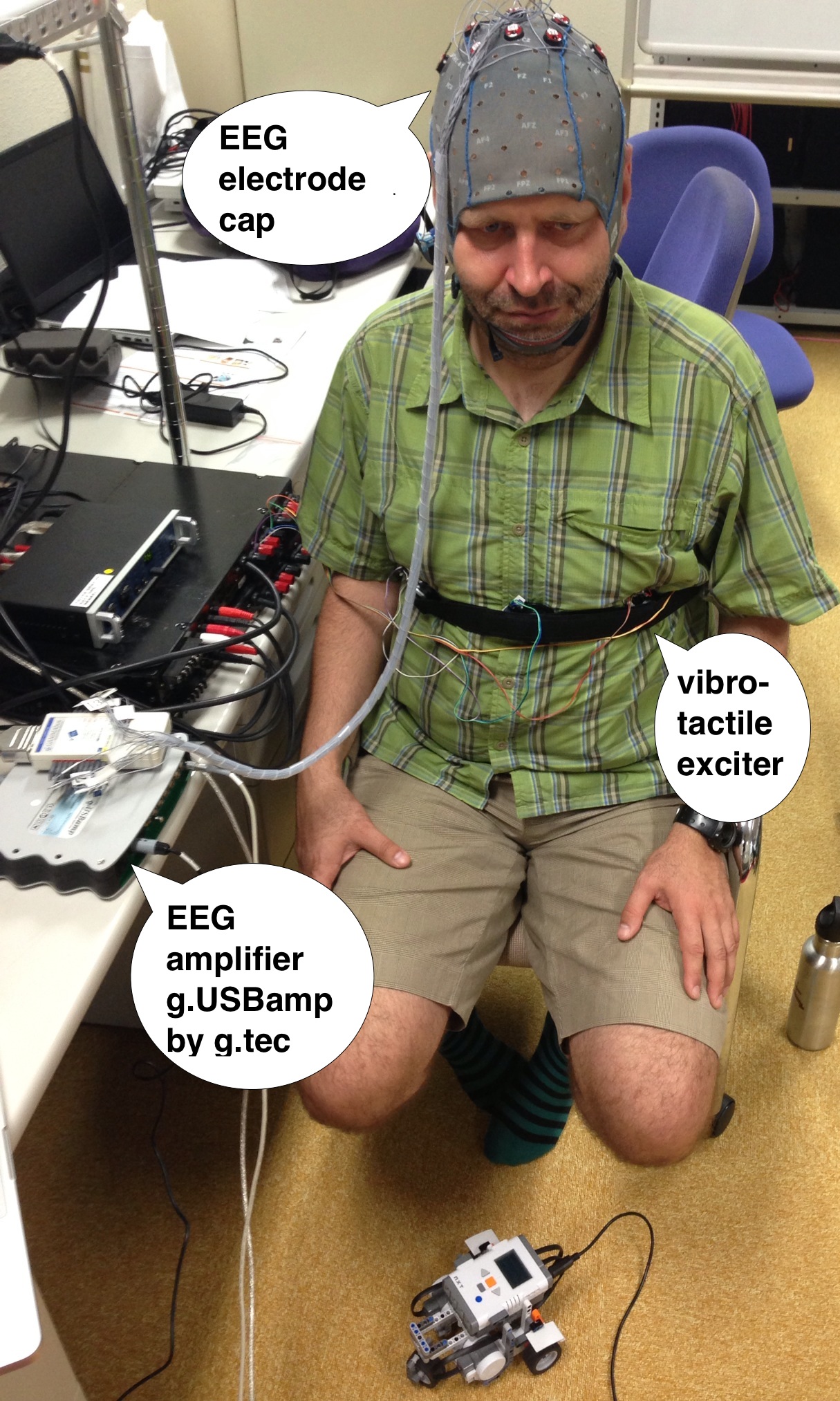}
	\end{center}
	\caption{Subject wearing EEG cap with $16$ active electrodes attached to the g.USBamp amplifier. The five vibrotactile transducers are attached to an elastic belt on the subject chest. Small vehicle robot by LEGO MINDSTROMS, operated via tBCI application developed by the authors, is placed on the floor in front of the subject in the picture.}\label{fig:subject}
\end{figure}

The tactile stimulus was delivered as sinusoidal wave generated by a portable computer with MAX/MSP software~\cite{maxMSP}. The stimuli were delivered via five channel outputs of an external \emph{digital-to-analog} signal converter RME~Fireface~UCX coupled with the two acoustic YAMAHA~P4050 power amplifiers (four acoustic frequency channels each). 
The stimuli were delivered to the subjects' chest locations via the tactile transducers HiWave HIAX25C10-8/HS operating in the acoustic frequency spectrum of $100-20,000$~Hz, as depicted in Figure~\ref{fig:subject}. Each transducer in the experiments was set to emit a sinusoidal wave at $200$~Hz to match the transducer's resonance frequency.  Tactile impulses were designed to stimulate the \emph{Pacini endings} (fast-adapting type II afferent type tactile sensory innervation receptors) which are the large receptive field mechanoreceptors in deeper layers of human skin~\cite{natureHAPTIC2009}.
The training instructions were presented visually by means of the \emph{BCI2000} program with numbers $1-5$ representing robot movement directions (see Table~\ref{tab:commands}) communicated via vibrotactile transducers attached to the subject's chest (see Figure~\ref{fig:subject}).

\subsection{EEG tBCI Experiment}
 
EEG signals were captured with an EEG amplifier system g.USBamp by g.tec Medical Engineering GmbH, Austria, using $16$ active electrodes. The electrodes were attached to the head locations: \emph{Cz, Pz, P3, P4, C3, C4, CP5, CP6, P1, P2, POz, C1, C2, FC1, FC2,} and \emph{FCz}, as in the $10/10$ extended international system~\cite{Jurcak20071600} (see the topographic plot in the top panel of Figure~\ref{fig:allTOPO}). The ground electrode was attached to \emph{FPz} position and reference to the left earlobe respectively. No electromagnetic interference was observed from the vibrotactile transducers operating in higher frequencies comparing to the EEG frequency spectrum. Details of the EEG experimental protocol are summarized in Table~\ref{tb:p4}.
The captured EEG signals were processed online by BCI2000-based application~\cite{bci2000book}, using a stepwise linear discriminant analysis (SWLDA) classifier~\cite{krusienski2006,p3gui} with features drawn from the $0-650$~ms ERP intervals. Additionally in offline mode the classification accuracy was compared and improved using linear SVM classifier~\cite{liblinear}. The EEG recording sampling rate was set at $512$~Hz, the high pass filter at $0.1$~Hz, and the low pass filter at $60$~Hz. The ISI was $400$~ms, and each stimulus duration was $100$~ms. The subjects were instructed to spell out the number sequences (corresponding to the interactive robotic application commands shown in Table~\ref{tab:commands}) communicated by the transducers in each session. Each $target$ was presented seven times in a single command trial. Each subject performed three experimental sessions (randomized $35~targets$ and $140~non-targets$ each), which were later averaged for the online SWLDA classifier case or treated as single trial (only the first ERP was used from a sequence of seven) for linear SVM, as discussed in a next section. The first online tBCI session was a subject practice, the second was used for training the classifiers, while the last experimental session was used for testing interfacing accuracy.

\section{Results}\label{sec:results}

This section discusses the results obtained in the EEG online and offline data processing experiments, which are summarized in in Table~\ref{tb:bpmr3} and Figures~\ref{fig:allTOPO},~\ref{fig:allERP},~\ref{fig:s3ERP},~and~\ref{fig:s2ERP}. All the participating in the study tBCI subjects scored well above the chance level of $20\%$, reaching an ITR in the range from $0.64$~bit/min to $5.14$~bit/min in case of the online BCI experiment with SWLDA classifier, which may be considered to be a good outcome for online vehicular robot driving experiments. The ITR was calculated as follows~\cite{bciSPATIALaudio2010}:
\begin{eqnarray}
	&ITR& = V \cdot R\\
	&R& = log_2 N + P\cdot log_2 P + (1-P)\cdot log_2\left(\frac{1-P}{N-1}\right),
\end{eqnarray}
where $R$ stands for the number of bits/selection; $N$ is the number of classes ($5$ in this study); $P$ is the classifier accuracy (see Table~\ref{tb:bpmr3}); and $V$ is the classification speed in selections/minute ($4.3$~selections/minute for this study in case of averaging of seven responses in online SWLDA case or $30$~selections/minute for the single trial based linear SVM classifier). The maximum ITR possible for the BCI subjects to achieve in the settings presented were $5.14$~bit/min and $69.65$~bit/min, for seven averaged and single trial cases respectively.

\section{Conclusions}

This case study demonstrated results obtained with a novel five--commands and chest locations based tBCI paradigm developed and used in experiments with five ``body--able'' subjects. The proposed interface could be used for a real--time operation of the robotic vehicle. 
The experiment results obtained in this study confirmed the validity of the chest tBCI for interactive applications and the possibility to further improve the results with utilization of the single trial based linear SVM classifier.

The EEG experiment with the paradigm has confirmed that tactile stimuli can be used to operate robotic devices with five commands and with the interfacing rate ranging from $0.64$~bit/min to $5.14$~bit/min for online case using SWLDA and $4.53$~bit/min to $69.95$~bit/min in the offline post processing case with linear SVM classifier respectively. 

The results presented offer a step forward in the development of novel neurotechnology applications. Due to the still not very high interfacing rate achieved in online BCI case among the subjects, the current paradigm would obviously need improvements and modifications to implement also online the proposed and tested offline linear SVM classifier based processing. These needs determine the major lines of study for future research. However, even in its current form, the proposed tBCI can be regarded as a practical solution for LIS patients (locked into their own bodies despite often intact cognitive functioning), who cannot use vision or auditory based interfaces due to sensory or other disabilities.

We plan to continue this line of the tactile BCI research in order to further optimize the signal processing and machine learning (classification) methods. Next we will test the paradigm with the LIS patients in need for BCI technology. 

\subsubsection*{Author contributions} 

HM, TMR: Performed the EEG experiments and analyzed the data. TMR: Conceived the concept of the spatial tactile BCI and designed the EEG experiments. SM:  Supported the project. HM, TMR: Wrote the paper. 

\subsubsection*{Acknowledgments} 

This research was supported in part by the Strategic Information and Communications R\&D Promotion Program no. 121803027 of The Ministry of Internal Affairs and Communication in Japan, and by KAKENHI, the Japan Society for the Promotion of Science, grant no. 12010738. We also acknowledge the technical support of YAMAHA Sound \& IT Development Division in Hamamatsu, Japan.

\newpage
\section*{Figure Legends}

\begin{description}
	\item[Figure~\ref{fig:subject}] Subject wearing EEG cap with $16$ active electrodes attached to the g.USBamp amplifier. The five vibrotactile transducers are attached to an elastic belt on the subject chest. Small vehicle robot by LEGO MINDSTROMS, operated via tBCI application developed by the authors, is placed on the floor in front of the subject in the picture.
	\item[Figure~\ref{fig:allTOPO}] All the five subjects grand mean averaged results of the chest stimulation EEG experiment. The left panel presents the head topographic plot of the $target$ versus $non-target$ area under the curve (AUC), a measure commonly used in machine learning intra--class discriminative analysis. ($AUC > 0.5$ is usually assumed to be confirmation of feature separability~\cite{book:pattRECO}). The top right panel presents the largest difference as obtained from the data displayed in the bottom panel. The topographic plot also depicts the electrode positions. The fact that all the electrodes received similar AUC values (red) supports the initial electrode placement. The second panel from the top presents averaged SEP responses to the $target$ stimuli (note the clear P300 response in the range of $350-600$~ms). The third panel presents averaged SEP responses to the $non-target$ stimuli (no P300 observed). Finally, the bottom panel presents the AUC of $target$ versus $non-target$ responses (again, P300 could easily be identified).
	\item[Figure~\ref{fig:allERP}] All five subjects grand mean averaged results for each electrode plotted separately. The red and blue lines depict targets and non--targets respectively together with standard error bars. The P300 related responses could be observed in the $350-650$~ms latencies.
		\item[Figure~\ref{fig:s3ERP}] Averaged results for each electrode separately of the subject~$\#3$ for whom $P300$ response was dominating leading for the best latency used in classification. The red and blue lines depict targets and non--targets respectively together with standard error bars.
		\item[Figure~\ref{fig:s2ERP}] Averaged results for each electrode separately of the subject~$\#2$ for whom $N200$ response was significant leading for improved early latency used in classification. Observe the very short latency around $200$ms where the standard error bars don't overlap. The red and blue lines depict targets and non--targets respectively together with standard error bars.
		
\end{description}

\clearpage

\section*{Tables}

\begin{table}[!h]
	\begin{center}
	\caption{Conditions of the EEG experiment.}\label{tb:p4}
	\begin{tabular}{|l|p{6cm}|}
	\hline
	Number of subjects					& $5$ \\
	\hline
	Tactile stimulus length			& $100$~ms \\
	\hline
	Stimulus frequency 		& $200$~Hz \\ 
	\hline
	Inter-stimulus-interval (ISI)		& $400$~ms \\
	\hline
	EEG recording system       			& \emph{g.USBamp} active EEG electrodes system.\\
	\hline
	Number of the EEG channels			& $16$\\
	\hline
	EEG electrode positions				& \emph{Cz, Pz, P3, P4, C3, C4, CP5, CP6, P1, P2, POz, C1, C2, FC1, FC2, FCz}\\
	\hline
	Reference and ground electrodes   	& earlobe and \emph{FPz} \\
	\hline
	Stimulus generation					& $5$ \emph{HIAX25C10-8} transducers\\
	\hline
	Number of trials used by SWLDA (SVM) 	& $7$ $(1)$ \\
	\hline	
	\end{tabular} 
	\end{center}
\end{table}

\clearpage

\begin{table}
	\begin{center}
	\caption{Interactive vehicular robot driving application commands encoded with the chest position numbers.}\label{tab:commands}
	\begin{tabular}{c|l}
	~~Chest position number~~ 	& ~~Command~~ \\
	\hline
	$1$				& ~~go left $(-90^\circ)$ \\
	$2$				& ~~go straight--left $(-45^\circ)$ \\
	$3$				& ~~go straight $(0^\circ)$ \\
	$4$				& ~~go straight--right $(45^\circ)$ \\
	$5$				& ~~go right $(90^\circ)$ \\
	\end{tabular} 
	\end{center}
	\vspace{-.5cm}
\end{table}

\clearpage

\begin{table}
	\begin{center}
	\caption{The chest positions stimulation EEG experiment accuracy and ITR scores. The theoretical chance level was $20\%$. For the SWLDA classifier, features were derived from the averages of the seven ERPs of all the subjects. In case of the linear SVM classifier only single trials (sequences) were used. Observe the increase in ITR with single trial classification in case of the linear SVM classifier utilization.}\label{tb:bpmr3}
	\begin{tabular}{| c | c | c | c |}
	\hline
	\multicolumn{4}{|c|}{\rule[-5pt]{0pt}{14pt} SWLDA classifier~\cite{krusienski2006,p3gui}}\\
	\hline \hline
	~Subject~  	& ~Number of averaged trials~ 	& ~Maximum accuracy~ 	& ITR \\
	\hline  
	%\hline
	$\#1$					& $7$					& $40\%$			& ~~$0.64$~bit/min~~ \\
	%\aline
	$\#2$					& $7$					& $60\%$			& ~~$2.36$~bit/min~~ \\ 
	%\hline
	$\#3$					& $7$					& $40\%$ 		& ~~$0.64$~bit/min~~ \\
	$\#4$					& $7$					& $80\%$ 		& ~~$5.14$~bit/min~~ \\
	$\#5$					& $7$					& $60\%$ 		& ~~$2.36$~bit/min~~ \\
	\hline \hline
	\multicolumn{4}{|c|}{\rule[-5pt]{0pt}{14pt} Offline linear SVM classifier~\cite{liblinear} based improvement}\\
	\hline \hline
	~Subject~  			& ~Number of averaged trials~ 	& ~Maximum accuracy~ 	& ITR \\
	\hline  
	%\hline
	$\#1$					& $1$					& $100\%$			& ~~$69.65$~bit/min~~ \\
	%\aline
	$\#2$					& $1$					& $60\%$			& ~~$16.53$~bit/min~~ \\ 
	%\hline
	$\#3$					& $1$					& $40\%$ 			& ~~$~4.53$~bit/min~~ \\
	$\#4$					& $1$					& $80\%$ 			& ~~$36.00$~bit/min~~ \\
	$\#5$					& $1$					& $60\%$ 			& ~~$16.53$~bit/min~~ \\
	\hline
	\end{tabular} 
	\end{center}
	\vspace{-0.4cm}
\end{table}

\clearpage

\section*{Figures}

\clearpage

\end{document}